\documentclass[aps,prb,twocolumn,showpacs,showkeys,groupedaddress,floatfix,amsmath,amssymb]{revtex4}
\usepackage{graphicx}
\usepackage{dcolumn}
\usepackage{bm}

\begin{document}


\title{Dynamic transitions between metastable states in a superconducting ring}

\author{D. Y. Vodolazov}
\author{F. M. Peeters}
\email{peeters@uia.ua.ac.be} \affiliation{Departement Natuurkunde,
Universiteit Antwerpen (UIA), Universiteitsplein 1,  B-2610
Antwerpen, Belgium}

\date{\today}

\begin{abstract}

Applying the time-dependent Ginzburg-Landau equations, transitions
between metastable states of a superconducting ring are
investigated in the presence of an external magnetic field. It is
shown that if the ring exhibits several metastable states at a
particular magnetic field, the transition from one metastable
state to another one is governed by {\it both} the relaxation time
of the absolute value of the order parameter $\tau_{|\psi|}$ and
the relaxation time of the phase of the order parameter
$\tau_{\phi}$. We found that the larger the ratio
$\tau_{|\psi|}/\tau_{\phi}$ the closer the final state will be to
the absolute minimum of the free energy, i.e. the thermodynamic
equilibrium. The transition to the final state occurs through a
subsequent set of {\it single} phase slips at a particular point
along the ring.

\end{abstract}


\maketitle

\section{Introduction}

The fast development of experimental techniques makes possible the
study of physical properties of samples with sizes of about
several nanometers. The interest in such small objects is due to
the appearance of new effects if the size of the system is
comparable to some characteristic length. In case of
superconductors it means that the sample should have one size at
least of order the coherence length $\xi$ (such superconductors
are called 'mesoscopic' superconductors). For example, only for
hollow cylinders or rings of radius $\sim \xi$ it is possible to
observe the famous Little-Parks oscillations \cite{Little}.

The majority of the research in this area has been limited to the
study of static or quasi-static properties of mesoscopic
superconducting rings, disks and other geometries \cite{Peeters}.
It is surprising that the dynamics in such systems was practically
not studied. However, it knows out that the dynamics is very
important for systems which exhibit a series of metastable states
and which may be brought far from thermal equilibrium. For such
systems the fundamental problem is to determine the final state to
which the system will transit too (see for example Ref.
\cite{Cross}).

In the present work, we present for the first time a detailed
study of the dynamic transitions between different states in a
mesoscopic superconducting ring. This system is a typical example
of the above mentioned systems where a set of metastable states
exist. We will show that the final state depends crucially on the
ratio between the relaxation time of the absolute value of the
order parameter and the time of change of the phase of the order
parameter. Our theoretical results explain the recent
magnetization results of Pedersen {\it et al.} \cite{Pedersen} in
thin and narrow superconducting $Al$ rings.

To solve this problem we present a numerical study of the
time-dependent Ginzburg-Landau (TDGL) equations. Therefore, our
results will also be applicable to other systems (for example,
liquid helium) where the dynamics are described by these or
similar equations. In our approach we neglect the self-field of
the ring (which is valid if the width and thickness of the ring
are less than $\lambda$ - the London penetration length) and hence
the distribution of the magnetic field and the vector potential
are known functions. The time-dependent GL equations in this case
are
\begin{subequations}
\begin{eqnarray}
u \left(\frac {\partial \psi}{\partial t}+i\varphi\psi  
\right) & = & (\nabla - {\rm i} {\bf A})^2 \psi
+\psi(1-|\psi|^2)+\chi, \quad
\\
\Delta \varphi & = &  {\rm div}\left({\rm Im}(\psi^*(\nabla-{\rm
i}{\bf A})\psi)\right), \quad
\end{eqnarray}
\end{subequations}
where $\psi=|\psi|e^{i\phi}$ is the order parameter, the vector
potential $A$ is scaled in units $\Phi_0/(2\pi\xi)$ (where
$\Phi_0$ is the quantum of magnetic flux), and the coordinates are
in units of the coherence length $\xi(T)$. In these units the
magnetic field is scaled by $H_{c2}$ and the current density, $j$,
by $j_0=c\Phi_0/8\pi^2\lambda^2\xi$. Time is scaled in units of
the Ginzburg-Landau relaxation time $\tau_{GL}=4\pi\sigma_n
\lambda^2/c^2 $, the electrostatic potential, $\varphi$, is in
units of $c\Phi_0/8 \pi^2 \xi \lambda \sigma_n$ ($\sigma_n $ is
the normal-state conductivity), and $u$ is a relaxation constant.
In our numerical calculations we used the new variable
$U=exp(-i\int {\bf A} d{\bf r})$ which guarantees gauge-invariance
of the vector potential on the grid. We also introduced small
white noise $\chi$ in our system, the size of which is much
smaller than the barrier height between the metastable states.

We consider $u$ as an adjustable parameter which is a measure of
the different relaxation times (for example, the relaxation time
of the absolute value of the order parameter) in the
superconductor. This approach is motivated by the following
observations. From an analysis of the general microscopic
equations, which are based on the BCS model, the relaxation
constant $u$ was determined in two limiting cases: $u=12$ for
dirty gapless superconductors \cite{Gor'kov} and $u=5.79$ for
superconductors with weak depairing \cite{Schmid,Kramer}. However,
this microscopic theory is built on several assumptions - for
example that the electron-electron interaction is of the BCS form,
which influences the exact value of $u$. On the other hand, the
stationary and time-dependent Ginzburg-Landau equations, Eqs.
(1a,b), are in some sense more general and their general form does
not depend on the specific microscopic model (but the value of the
parameters, of course, are determined by the microscopic theories)
\cite{self1}.

The paper is organized as follows. In Sect. II we present our
numerical results for the solution of Eqs. (1a,b) for strictly
one-dimensional ring. In Sect. III we explain the obtained results
in terms of different time scales of the mesoscopic
superconductor. The effect of the finite width of the ring is
studied in Sect. IV where we also discuss the influence of
non-zero temperature and finite $\lambda$.

\section{Superconducting ring out of equilibrium}

If the width of the ring is much smaller than $\xi$ and
$\lambda_{\bot}=\lambda^2/d$ ($d$ is the film thickness) it is
possible to consider the ring, of radius $R$, as a one-dimensional
object. At first we limit ourselves to the solution of the
strictly one-dimensional Eqs. (1a,b) which model the dynamics of a
ring with small width. In our approach the vector potential is
equal to $A=HR/2$, where $H$ is the applied magnetic field.
\begin{figure}[h]
\includegraphics[width=0.5\textwidth]{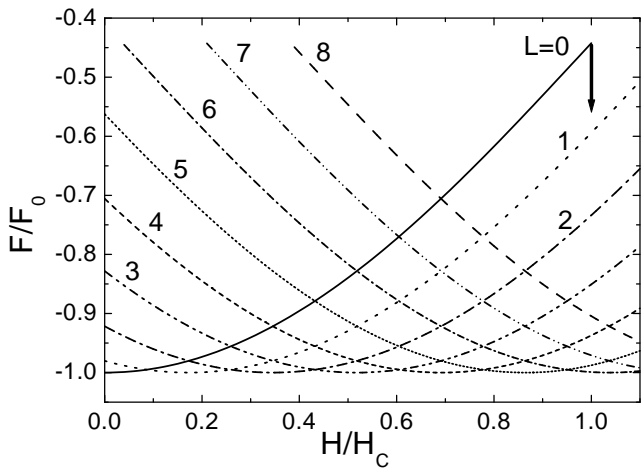}
\caption{Dependence of the free energy of the one-dimensional ring
(with $R=10\xi$) on the applied magnetic field for different
vorticity $L$.}
\end{figure}

As shown in Refs. \cite{Tarlie,Vodol1} the transition of the
superconducting ring from a state with vorticity $L=\oint \nabla
\phi ds/2\pi$ (which in general can be a metastable state) to a
state with a different vorticity occurs when the absolute value of
the gauge-invariant momentum ${\bf p}=\nabla \phi-{\bf A}$ reaches
the critical value
\begin{equation}
p_c=\frac{1}{\sqrt{3}}\sqrt{1+\frac{1}{2R^2}}.  
\end{equation}
Let us, for simplicity, consider that the magnetic field is
increased from zero (with initial vorticity of the ring $L=0$) to
the critical $H_c$ where $p$ becomes equal to $p_c$ (for $L=0$ we
have $H_c=2p_c/R$). In this case ${\bf p}=-{\bf A}$ and we can
write Eq. (2) as
\begin{equation}
\Phi_c/\Phi_0=\frac{R}{\sqrt{3}}\sqrt{1+\frac{1}{2R^2}}. 
\end{equation}
For this value of the flux the thermodynamical equilibrium state
becomes $L_{eq}={\rm Int}(\Phi_c/\Phi_0)$. For example, if $R=10$
we find $L_{eq}=6$ (see Fig. 1). The fundamental question we want
to answer is: {\it what will be the actual value of the vorticity
of the final state}? Will it be the thermodynamic equilibrium
state or a metastable one? This answer will be obtained from a
numerical solution of the time-dependent Ginzburg-Landau
equations.
\begin{figure}[h]
\includegraphics[width=0.5\textwidth]{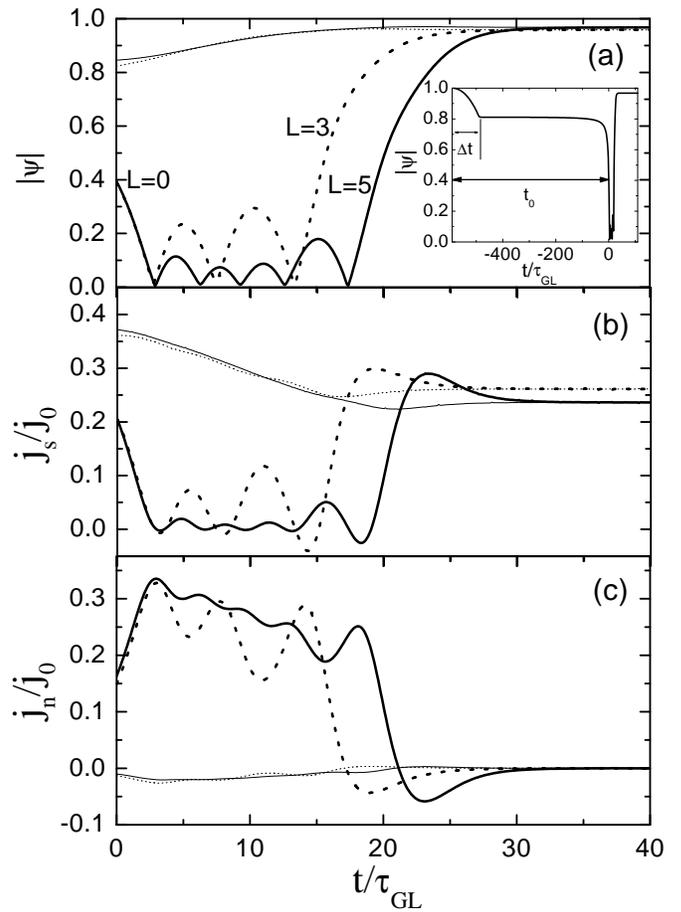}
\caption{Dependence of the order parameter (a), the
superconducting $j_s$ (b) and the normal $j_n$ (c) current density
in the point where the minimal (thick curves) and the maximal
(thin curves) values of the order parameter is reached as function
of time for $u=3$, $R=10$ (dotted curves) and $R=15$ (solid
curves). For both rings we took $\Delta t=100$, the time interval
over which the magnetic field is increased to its value $H_c$
(inset of Fig. 2(a)).}
\end{figure}

For our numerical calculations we considered two different rings
with radius $R=10$ and $R=15$. These values were chosen such that
the final state exhibits several metastable states and, in
principle, transitions can occur with jumps in $L$ which are
larger than unity. For $R=10$, $\Delta L$ may attain values
between 1 and 6 and for $R=15$ it may range from 1 to 9. Smaller
rings have a smaller range of possible $\Delta L$ values. We found
that larger rings than $R=15$ did not lead to new effects and are
therefore not considered here. In our numerical simulations we
increased $H$ gradually from zero to a value $H_c +\Delta H$ (we
took $\Delta H=0.036H_c$ for $R=10$ and $\Delta H=0.012 H_c$ for
$R=15$) during a time interval $\Delta t$, after which the
magnetic field was kept constant. The magnetic field $\Delta H$
and time $\Delta t$ range was chosen sufficiently large in order
to speed up the initial time for the nucleation of the phase slip
process, but still sufficiently small in order to model real
experimental situations in which $H$ is increased during a time
much larger than $u\cdot \tau_{GL}$ (for example for $Al$
$\tau_{GL}(T=0)\simeq 10^{-11}s$). A change of $\Delta H$ and
$\Delta t$, within realistic boundaries, did not have an influence
on our final results.

From our detailed numerical analysis, we found that the vorticity
$L$ of the final state of the ring depends on the value of $u$ and
is not necessary equal to $L_{eq}$. The larger $u$ the larger the
vorticity after the transition. The final state is reached in the
following manner. When the magnetic field increases the order
parameter decreases (see inset of Fig. 2(a)). First, in a single
point of the ring a local suppression, in comparison with other
points, of the order parameter occurs which deepens gradually with
time during the initial part of the development of the
instability. This time scale is taken as the time in which the
value of the order parameter decreases from 1 to 0.4 in its
minimal point. When the order parameter reaches the value $\sim
0.4$ in its minimal point the process speeds up considerably and
the order parameter starts to oscillate in time at this point
along the ring (see Fig. 2(a)). At the same time also oscillatory
behavior of the superconducting $j_s={\rm Re}(\psi^*(-{\rm
i}\nabla-{\bf A})\psi)$ (Fig. 2(b)) and the normal $j_n=-\nabla
\varphi$ (Fig. 2(c)) current density is found at the point where
the minimal value of the order parameter is found. At other places
in the ring such oscillatory behavior is strongly damped (see e.g.
the thin curves in Fig. 2). After some time the system evolves to
a new stable state, which in the situation of Fig. 2 is $L=3$ for
$R=10$ and $L=5$ for $R=15$.
\begin{figure}[h]
\includegraphics[width=0.5\textwidth]{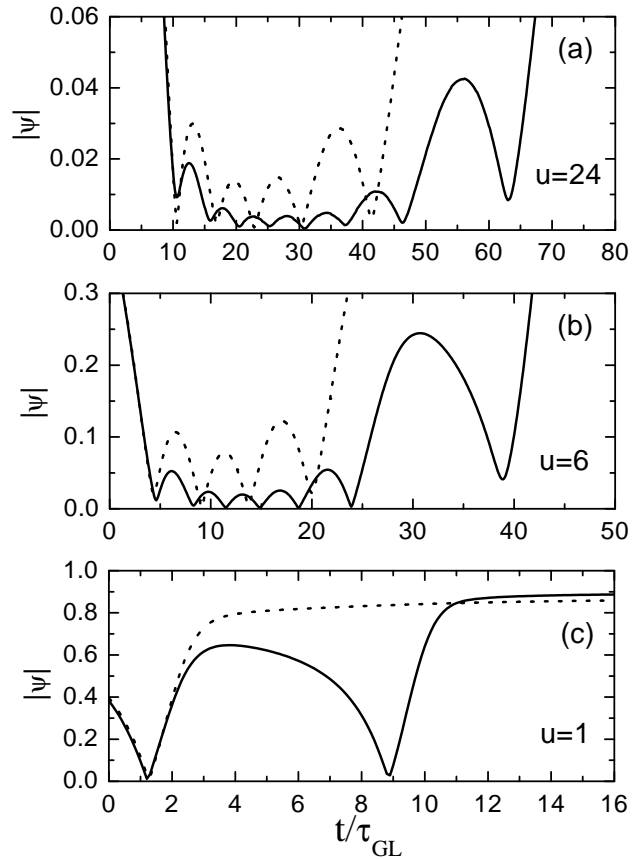}
\caption{The dynamics of the order parameter in its minimal point
for different values of the parameter $u$ and for two values of
the radius $R=10\xi$ (dotted curves) and $R=15\xi$ (solid curves).
The zero of time is taken at the moment when the value of the
order parameter becomes equal to $0.4$ in its minimal point.
Nonzero values of $|\psi|$ near the moment of the phase slip is
connected with the finite coordinate and time step used in the
numerical calculations.}
\end{figure}

In Fig. 3 the dependence on $u$ of the dynamics of $|\psi|$ in the
minimal point is illustrated. Note that with increasing value of
$u$ the number of phase slips (or equivalently the number of
oscillations of the order parameter at its minimal point)
increases and hence $\Delta L$ also increases. Another important
property is that the amplitude of those oscillations decreases
with increasing $u$.

\section{Time scales}

The above results lead us to conclude that in a supeconducting
ring there are two characteristics time scales. First, there is
the relaxation time of the absolute value of the order parameter
$\tau_{|\psi|}$. The second time scale is determined by the time
between the phase slips (PS), i.e. the period of oscillation of
the value of the order parameter in its minimal point (see Fig.
3). Below we show that the latter time is directly proportional to
the time of change of the phase of the order parameter
$\tau_{\phi}$ and is connected to the relaxation time of the
charge imbalance in the system.

In Fig. 4 the distribution of the absolute value of the order
parameter $\psi$ and the gauge-invariant momentum $p$ are shown
near the point where the first phase-slip occurs for a ring with
radius $R=10$ and $u=3$ at different times: just before and after
the first PS which occurs at $t\simeq 2.7\tau_{GL}$. Before the
moment of the phase slip the order parameter decreases while after
the PS it increases. In order to understand this different
behavior let us rewrite Eq. (1a) separately for the absolute value
$|\psi|$ and the phase $\phi$ of the order parameter
\begin{subequations}
\begin{eqnarray}
u\frac {\partial |\psi|}{\partial t} &  
 = &  \frac{\partial^2 |\psi|}{\partial s^2}+|\psi|(1-|\psi|^2-p^2),
\qquad
\\
\frac {\partial \phi}{\partial t} & = &
-\varphi-\frac{1}{u|\psi|^2}\frac{\partial j_n}{\partial s} .
\qquad 
\end{eqnarray}
\end{subequations}
Here $s$ is the arc-coordinate along the ring and we used the
condition ${\rm div}(j_s+j_n)=0$. It is obvious from Eq. (4a) that
if the RHS of Eq. (4a) is negative $|\psi|$ decreases in time and
if the RHS is positive $|\psi|$ will increase in time. Because the
second derivative of $|\psi|$ is always positive (at least near
the phase-slip center - see Fig. 4(a)) the different time
dependence of $|\psi|$ is governed by the term $-p^2|\psi|$. From
Fig. 4(b) it is clear that after the phase slip the value of $p$
is less than before this moment, with practically the same
distribution of $|\psi|$. It is this fact which is responsible for
an increase of the order parameter just after the moment of the
phase slip. But at some moment of time the momentum $p$ can become
sufficiently large, making the RHS of Eq. (4a) negative and as a
consequence $|\psi|$ starts to decrease.
\begin{figure}[hbtp]
\includegraphics[width=0.48\textwidth]{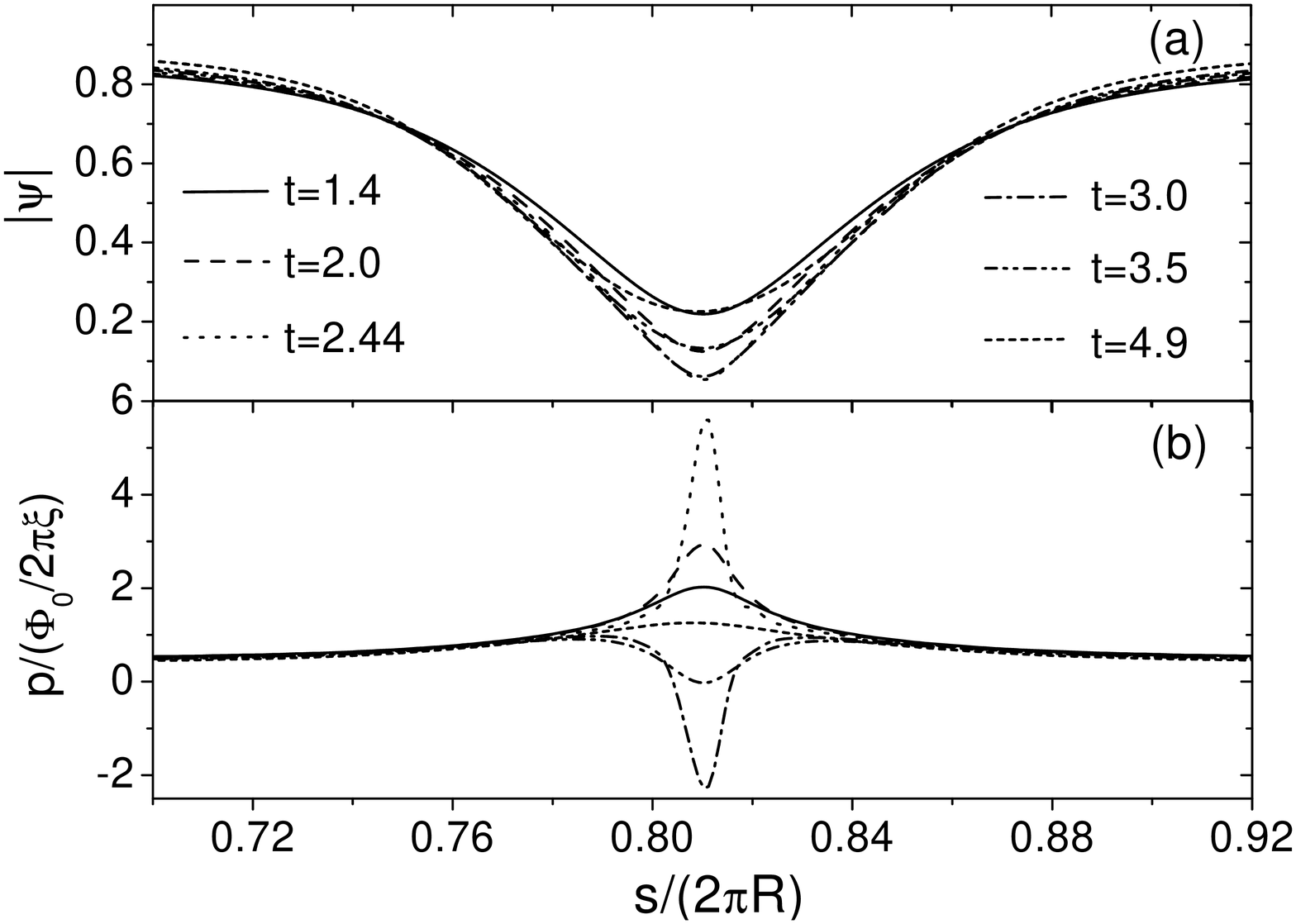}
\caption{Distribution of the absolute value of the order parameter
(a) and gauge-invariant momentum (b) near the phase slip center at
different moments of time for a ring of $R=10\xi$ and with $u=3$.
The phase slip occurs at $t\simeq 2.7\tau_{GL}$.}
\end{figure}

Based on our numerical calculations we may state that for every
value of the order parameter in the minimal point, there exists a
critical value of the momentum $p^{min}_c$ such that if the value
of the momentum in this point is less than $p^{min}_c$ the order
parameter will increase in time. In the opposite case the order
parameter decreases which leads ultimately to the phase slip
process. Therefore, when after a phase slip $p$ increases fast
enough such that at some moment the condition $p^{min}>p^{min}_c$
is fulfilled, the order parameter will start to decrease which
leads than to a new phase slip process.

From our results we can conclude that the change of $p$ (or phase
of the order parameter) with time and of $|\psi|$ with time has a
different dependence on $u$. Indeed from Figs. 2 and 3 it follows
that with increasing $u$ the time relaxation of $|\psi|$ becomes
larger with respect to the relaxation time for $p$ (for example,
the amplitude of oscillations of $|\psi|$ is decreased). Moreover,
the relaxation time for $p$ depends not only on $u$ but also on
the history of the system: the larger the number of phase slips
which have occurred in the system the longer the time becomes
between the next two subsequent phase slips. In order to
understand such behavior we turn to Eq. (4b).

The numerical analysis shows that the second term in the RHS of
Eq. (4b) is only important for some length near the phase slip
center. This length is nothing else than the length over which the
electric field penetrates the superconductor. We checked this
directly by solving Eqs. (1a,b) for this situation. But this
length is the decay length $\lambda_Q$ of the charge imbalance in
the superconductor (see for example Refs. \cite{Kadin,Tinkham}).
We numerically found that $\lambda_Q$ varies with $u$ (in the
range $u=1-100$) as $\lambda_Q \sim u^{-0.27}$. It means that the
relaxation time of the charge imbalance is $\tau_Q\sim
\lambda_Q^2\sim u^{-0.54}$.

Lets take the derivative $\partial/\partial s$ on both side of Eq.
(4b) and integrate it over a distance of $\lambda_Q$ near the PS
center (far from it we have $j_n \sim 0$) we obtain the Josephson
relation (in dimensionless units)
\begin{equation}
\frac{d \Delta \phi}{dt}=V \sim j_n(0)\lambda_Q, 
\end{equation}
where $\Delta \phi=\phi(+\lambda_Q)-\phi(-\lambda_Q)$ is the phase
difference over the phase slip center which leads to the voltage
$V=-(\varphi(+\lambda_Q)-\varphi(-\lambda_Q))$ and where $j_n(0)$
is the normal current (or electric field in our units) in the
point of the phase slip. From Eq. (5) it follows immediately that
the relaxation time for the phase of the order parameter near the
phase slip center is
\begin{equation}
\tau_{\phi} \sim \frac{1}{\lambda_Q <|j_n(0)|>}, 
\end{equation}
where $<\cdot>$ means averaging over time between two consecutive
phase slips. This result allows us to qualitatively explain our
numerical results. Indeed, from Fig. 2(c) it is apparent that
$<|j_n(0)|>$ decreases after each PS and as a result it leads to
an increase of $\tau_{\phi}$. With increasing $u$, $\lambda_Q$
decreases and $\tau_{\phi}$ increases. Fitting our data (see Fig.
5) leads to the dependencies $\tau_{\phi} \sim u^{0.21}$ and
$\tau_{\phi} \sim u^{0.23}$ for rings with radius $R=10$ and
$R=15$, respectively (here $\tau_{\phi}$ was defined as the time
between the first and the second phase slip). This is close to the
expected dependencies which follow from Eq. (6) and from the above
dependence of $\lambda_Q(u)$ (quantitative differences follows
from uncertainty in our finding $\tau_{\phi}$ and $\lambda_Q$).
Besides, Eq. (6) allows us to explain the decrease of
$\tau_{\phi}$ with increasing radius of the ring (see inset in
Fig. 5). Namely, during the time between two phase slips the
gauge-invariant momentum decreases as $\Delta p \sim 1/R$ (because
$\nabla \phi$ increases as $\sim 2\pi/2\pi R$) in the system. Far
from the phase slip center the total current practically is equal
to $j_s\sim p$. Because ${\rm div}(j_s+j_n)=0$ in the ring we
directly obtain that during this time the normal current density
in the point of the phase slip also decreases as $\sim 1/R$.
Taking into account Eq. (6) we can conclude that $\tau_{\phi}$
should vary as $\sim 1/R$ (at least for a large radius and for the
first phase slip). The behavior shown in the inset of Fig. 5 is
very close to such a dependence (it is interesting to note that in
contrast to $\tau_{\phi}$ the time $\tau_{|\psi|}$ does
practically not depend on $R$).

On the basis of our results we can make the following statement:
when the period of oscillation (time of change of the phase of the
order parameter) becomes of the order, or larger, than the
relaxation time of the absolute value of the order parameter the
next phase slip becomes impossible in the system.

This result can be applied to the system of Ref. \cite{Kramer2}
where a current carrying wire was studied. The authors found a
critical current $j_c=0.335$ for $u=5.79$ and $j_c=0.291$ for
$u=12$. From our observation follows that the phase slip solution
may be realized as a stable solution when
$\tau_{|\psi|}>\tau_{\phi}\sim 1/(\lambda_{Q} <|j_n(0)|>)$. We
found that $\lambda_Q \sim u^{-0.27}$, $\tau_{|\psi|}\sim u^{0.6}$
and $<|j_n(0)|> \sim j_{ext}$ which leads to the critical current
$j_c \sim u^{-0.33}$ which decreases with increasing $u$
\cite{comm1}.

Above we found that the time scale governing the change in the
phase does not coincide with the relaxation time of the absolute
value of the order parameter. This difference is essentially
connected with the presence of an electrostatic potential in the
system. In order to demonstrate this we performed the following
numerical experiment. We neglected $\varphi$ in Eq. (1a) and found
that the number of PS is now independent of $u$ and $R$, and
$\Delta L$ equals unity. Moreover, it turned out that the time
scale $t_0$ is an order of magnitude larger. This clearly shows
that the electrostatic potential is responsible for the appearance
of a second characteristic time which results in the above
mentioned effects.

In an earlier paper \cite{Tarlie} the question of the selection of
the metastable state was already discussed for the case of a
superconducting ring. However, these authors neglected the effect
of the electrostatic potential and found that transitions with
$\Delta L>1$ can occur only when the magnetic field (called the
induced electro-moving force (emf) in Ref. \cite{Tarlie})
increases very quickly. In their case these transitions were
connected to the appearance of several nodes in $\psi$ along the
perimeter of the ring and in each node a single PS occurred. We
reproduced those results and found such transitions also for
larger values of $\Delta H$. However, simple estimates show that
in order to realize such a situation in practice it is necessary
to have an extremely large ramp of the magnetic field. For
example, for $Al$ mesoscopic samples with $\xi(0)=100nm$ and
$R=10\xi$ the corresponding ramp should be about $10^3-10^4 T/s$.
With such a ramp the induced normal currents in the ring are so
large that heating effects will suppress superconductivity.

\begin{figure}[h]
\includegraphics[width=0.48\textwidth]{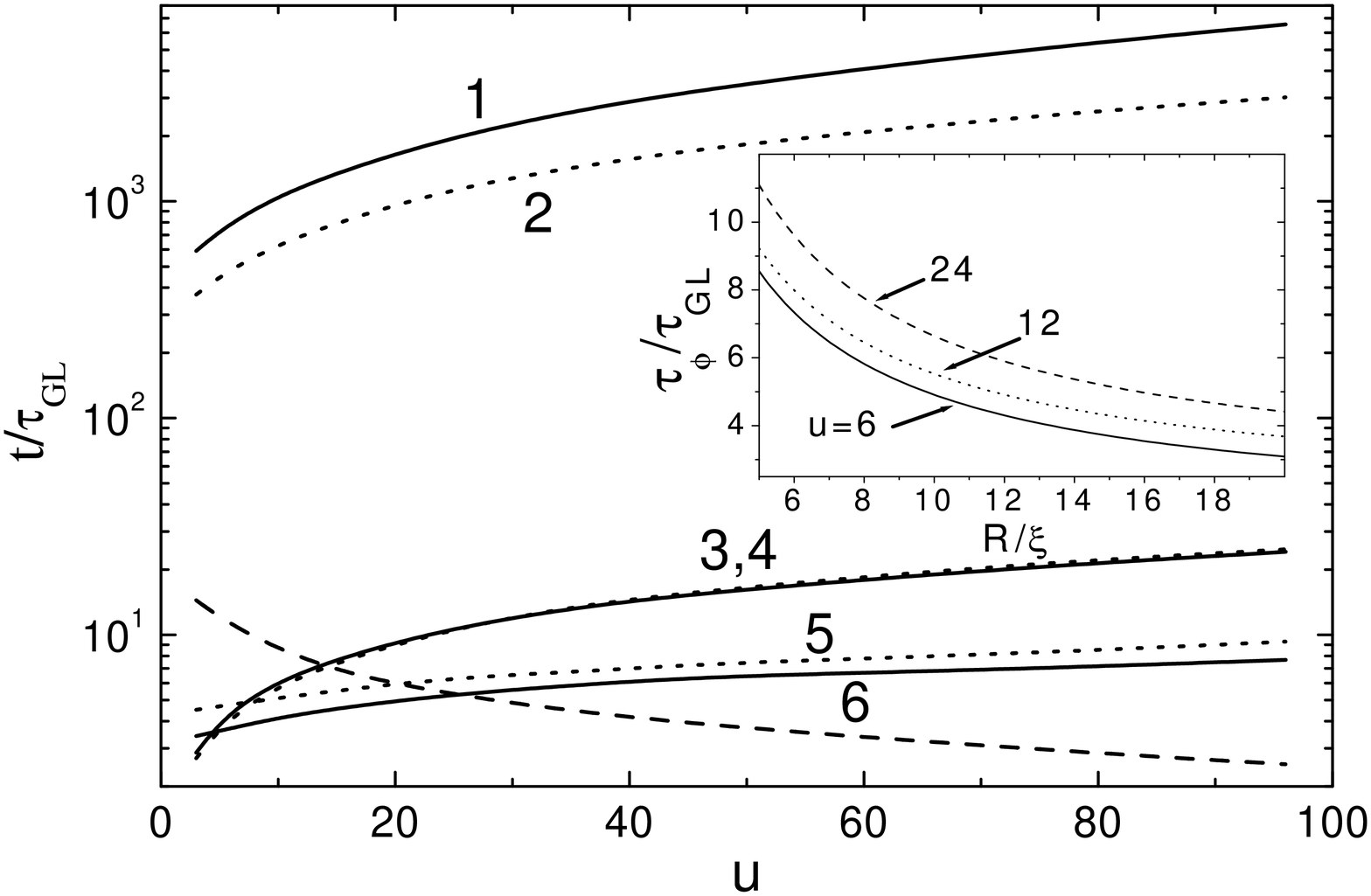}
\caption{Dependence of the initial nucleation time $t_0$ (curves
1,2), the relaxation time of the order parameter $\tau_{|\psi|}$,
(which for definiteness we defined as the time of variation of the
order parameter from $0.4$ till the first PS - curves 3,4) and the
relaxation time of the phase of the order parameter $\tau_{\phi}$
(we defined it here as the time between the first and second PS -
curves 5,6) are shown as function of the parameter $u$.
Dotted(solid) curves are for a ring with radius $R=10(15)$. Dashed
curve corresponds to the relaxation time of the charge imbalance
(as obtained from the expression $\tau_Q=5.79\lambda_Q^2$). In the
inset, the dependency of $\tau_{\phi}$ on the ring radius is shown
for different values of the parameter $u$.}
\end{figure}

The transitions between the metastable states in the ring are not
only determined by $u$ and $R$, but also, the presence of defects
play a crucial role. Their existence leads to a decrease of
$\Delta L$. This is mainly connected to the fact that with
decreasing $p_c$ (as a result of the presence of a defect) the
value of $<|j_n(0)|>$ decreases even at the moment of the first
phase slip (because $<|j_n(0)|>$ cannot be larger than
$<|j_s(0)|>\simeq p_c$) and hence $\tau_{\phi}$ increases. We can
also explain this in terms of a decreasing degeneracy of the
system. For example, if we decrease $p_c$ with a factor of two
(e.g. by the presence of defects) it leads to a twice smaller
value for $\Phi_c$ and $L_{eq}$. But our numerical analysis showed
that the effect of defects is not only restricted to a decrease of
$p_c$ and $L_{eq}$. To show this we simulated a defect by
inserting an inclusion of another material with less or zero
$T_c$. This is done by inserting in the RHS of Eq. (1a) an
additional term $\rho(r)\psi$ where $\rho(r)$ is zero except
inside the defect where $\rho(r)=\alpha<0$. The magnetic field was
increased up to $H_c$ of the ideal ring case. We found that the
number of PS was smaller as compared to the case of a ring without
defects. The calculations was done for defects such that $p_c$ was
decreased by less than a factor of 2 in comparison to the ideal
ring case. In contrast, a similar calculation for a ring with
nonuniform thickness/width showed that, even for 'weak'
nonuniformity (which decreases $p_c$ by less than $20\%$), $\Delta
L$ was larger than for the ideal ring case and the final vorticity
approaches $L_{eq}$. This remarkable difference between the
situation for a defect and the case of a nonuniformity may be
traced back to the difference in the distribution of the order
parameter: even in the absence of an external magnetic field the
defect leads to a nonuniform distribution of the order parameter
which is not so for the nonuniform ring case. A more thorough
study of the effect of defects will be presented elsewhere.

\section{Superconducting ring with nonzero width}

All the above results were based on a one-dimensional model which
contains the essential physics of the decay and recovery of the
superconducting state in a ring from a metastable state to its
final state. However, even in the case when the effect of the
self-field can be neglected the finite width of the ring may still
lead to important additional effects (for example, a finite
critical magnetic field). In order to include the finite width of
the ring into our calculation we considered the following model.
We took a ring of mean radius $R=12\xi$, width $3.5\xi$, and
thickness less than $\xi$ and $\lambda$. These parameters are
close to those of the ring studied experimentally in Ref.
\cite{Pedersen} and they are such that we can still neglect the
self-field of the ring. The obtained magnetization curves of such
a ring are shown in Fig. 6 as function of the magnetic field and
for two values of $u$. Those results were obtained from a
numerical solution of the two-dimensional Ginzburg-Landau
equations, Eqs. (1a,b). The magnetic field was changed with steps
$\Delta H$ over a time interval which is larger than the initial
part of the phase slip process $t_0$ (for our parameters this
procedure leads practically to an adiabatic change of the magnetic
field).

From Fig. 6 we notice that the value of the vorticity jumps,
$\Delta L$, depends sensitively on the parameter $u$ \cite{comm2}.
We found that the phase slips occur in one particular place along
the perimeter of the ring. However, in contrast to our previous
one-dimensional case, $\Delta L$ depends also on the applied
magnetic field. The reason is that for a finite width ring the
number of metastable states decreases with increasing magnetic
field (see Ref. \cite{Berger}). It means that the system cannot be
moved far from equilibrium with a large superconducting current
density (because the order parameter is strongly suppressed by the
external field) at high magnetic field. Hence, the value of
$<|j_n(0)|>$ will be much smaller in comparison with the one at
low magnetic fields and $\tau_{\phi}$ is larger or comparable with
$\tau_{|\psi|}$ even for high values of $u$ and even for the first
PS. Thus, the effect of a large magnetic field in case of a finite
width ring is similar, to some respect, to the effect of defects
for a one-dimensional ring.

Our numerical results are in {\it qualitative} agreement with the
experimental results of Pedersen {\it et al.} (see Fig. 2 of Ref.
\cite{Pedersen}). Unfortunately, no {\it quantitative} comparison
is possible because of a number of unknowns, e.g. the value of
$p_c$ (it is necessary to have the dependence of $M(H)$ as
obtained starting from zero magnetic field), the value of $\xi$ is
not accurately known and hence the ratio $R/\xi$ can therefore
only be estimated. The value of both these parameters have a
strong influence on the value of the vorticity jumps $\Delta L$.
\begin{figure}[h]
\includegraphics[width=0.48\textwidth]{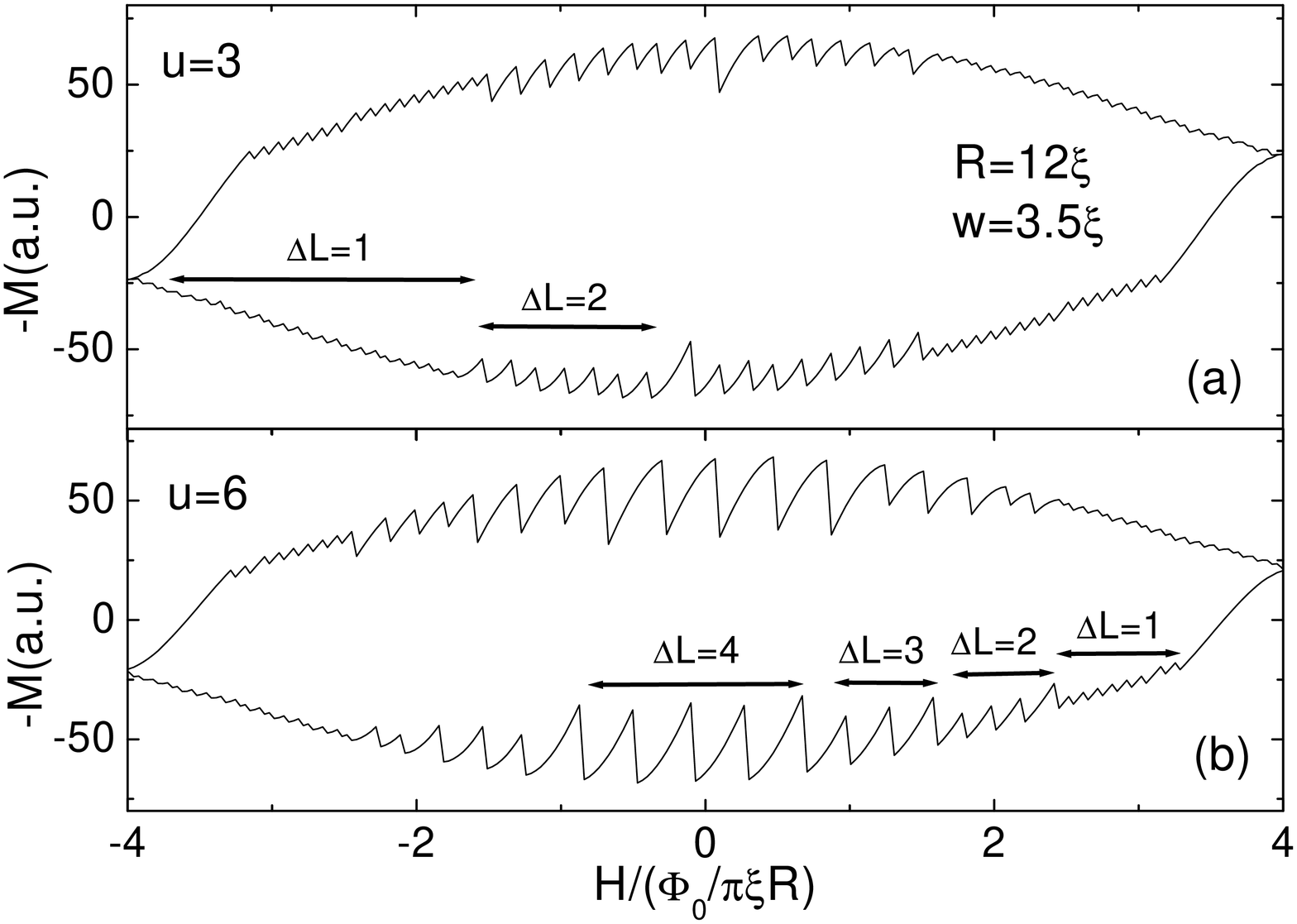}
\caption{Dependence of the magnetization of a finite width ring on
the external magnetic field for two values of the parameter $u$.
The results are shown for a sweep up and sweep down of the
magnetic field.}
\end{figure}

Let us discuss now finite temperature and screening effects (i.e.
finite $\lambda$). Nonzero temperature leads to the possibility of
a perturbative appearance of phase slip centers at $H^*<H_c$. But
if the field $H^*$ does not differ too much from $H_c$ the average
$<|j_n(0)|>$ is not altered strongly. As a result, after the
appearance of the first thermo-activated PS center the second
phase slip may appear automatically (if the ratio
$\tau_{|\psi|}/\tau_{\phi}$ is large enough) which leads to an
avalanche-type of process. Unfortunately, it is impossible to
apply the results of Ref. \cite{McCumber} in order to calculate
the probability of this process even for the first PS. The reason
is that in Ref. \cite{McCumber} the effect of the electrostatic
potential (which play a crucial role as was mentioned above) was
neglected.

A finite $\lambda$ (and large enough width of the ring) leads to a
nonuniform distribution of the momentum $p$ over the width of the
ring which can play an important role on the above considered
effects. As we mentioned above such a finite width ring may be
modelled as a strip with transport current in an external magnetic
field. In Ref. \cite{Aranson} the condition for the entry of the
first vortices in a narrow superconducting strip was studied. It
turned out, that the period of vortex chain entry depends
essentially on the distribution of $p$ over the width of the
strip. When this distribution is uniform the period is infinite.
But even small nonuniformities in this distribution leads to a
finite period. If we translate their results to our ring geometry,
the period of the vortex chain must now be discrete $2\pi R/n$
($n=1,2,...$). Under a certain condition the entry of vortices
becomes possible, not through a single point along the perimeter,
but through two, three and more points (without increasing the
ramp of the magnetic field). A finite $\lambda$ only increases
this effect because it leads to larger nonuniformities in the
dependence of $p$ over the ring width. In this situation there is
a competition between the process of the appearance of additional
nodes along the ring perimeter and the number of PS in these
points. At high magnetic fields these effects becomes negligible
small (because superconductivity only exists near the edges and
the ring effectively can be considered as two one-dimensional
rings) and we will have the situation as discussed in the present
work, and, recently in \cite{Berger}.

\section{Conclusion}

In conclusion, we studied how an unstable superconducting state of
a superconducting ring evolves in time and transit to its final
state. The latter is not necessary the thermodynamic equilibrium
state and may be another metastable state with a different
vorticity but which is stable in time. The transition between the
different superconducting states occurs through a phase slip
center which is a point along the ring where the superconducting
amplitude decreases to zero abruptly resulting in the change of
the vorticity of the superconducting state with one unit. The
waiting time, or the creation time for the first phase slip, is
found to be two orders of magnitude larger than the subsequent
time intervals between consecutive phase slips. The latter time is
connected with the time relaxation of the charge imbalance in the
superconductor and increases the closer the system becomes to the
final state. This circumstance allows to find this time (also as
the relaxation time of the absolute value of the order parameter)
from magnetic measurements of superconducting rings of large
radius. Our theoretical findings are in agreement with recent
experimental results of Pedersen {\it et al.} \cite{Pedersen}.

\begin{acknowledgments}
The work was supported by the Flemish Science Foundation (FWO-Vl),
the "Onderzoeksraad van de Universiteit Antwerpen", the
"Interuniversity Poles of Attraction Program - Belgian State,
Prime Minister's Office - Federal Office for Scientific, Technical
and Cultural Affairs," and the European ESF-Vortex Matter. One of
us (D.Y.V.) is supported by a postdoctoral fellowship of FWO-Vl.
\end{acknowledgments}


\begin{references}
\bibitem{self1} For a discussion of this question see also Refs. \cite
{Rieger,Dorsey} and for a review of different types of
time-dependent Ginzburg-Landau equations we refer to Refs.
\cite{Ivlev,Lozanne,Kadin,Tinkham}.
\bibitem{comm2} The monotonic change of the magnetization $M$ at the switching
magnetic field, i.e. $|H/(\Phi_0/\pi\xi R)|$, from increasing to
decreasing behavior is connected with the conservation of
vorticity $L$ of the ring in this range of magnetic fields.
\end{references}
\end{document}